# Simultaneous Identification and Control Using Active Signal Injection for Series Hybrid Electric Vehicles based on Dynamic Programming

Haojie Zhu[a], Ziyou Song[b*], Jun Hou[c*], Heath Hofmann[c], Jing Sun[b]

[a]Department of Mechanical Engineering, University of Michigan, Ann Arbor, MI 48109, USA

[b]Department of Naval Architecture and Marine Engineering, University of Michigan, Ann Arbor, MI 48109, USA

[c]Department of Electrical Engineering and Computer Science, University of Michigan, Ann Arbor, MI 48109, USA

**Abstract—**Hybrid electric vehicles (HEVs) have an over-actuated system by including two power sources, a battery pack and an internal combustion engine. This feature of HEV is exploited in this paper to simultaneously achieve accurate identification of battery parameters/states. By actively injecting current signals, state of charge, state of health, and other battery parameters can be estimated in a specific sequence to improve the identification performance when compared to the case where all parameters and states are estimated concurrently using the baseline current signals. A dynamic programming strategy is developed to provide the benchmark results about how to balance the conflicting objectives corresponding to identification and system efficiency. The tradeoff between different objectives is presented to optimize the current profile so that the richness of signal can be ensured and the fuel economy can be optimized. In addition, simulation results show that the Root-Mean-Square error of the estimation can be decreased by up to 100% at a cost of less than 2% increase in fuel consumption. With the proposed simultaneous identification and control algorithm, the parameters/states of the battery can be monitored to ensure safe and efficient application of the battery for HEVs.

**Nomenclature**

| | |
|---|---|
| $a$ | Linearized OCV-SOC slope (V) |
| $A_{front}$ | Frontal area of the HEV ($m^2$) |
| $b$ | Constants of linearized OCV-SOC curve (V) |
| $C_D$ | Air drag coefficient |
| $C_t$ | Capacitance of the RC pair (F) |
| $f$ | Rolling resistance coefficient |
| $f_{injected}$ | Frequency of the injected signal (Hz) |
| $g$ | Gravitational acceleration (m/s$^2$) |
| $i_b$ | Battery current (A) |
| $i_{bf}$ | Filtered terminal current of the battery (A) |
| $i_{b,max}$ | Upper bound of the battery current (A) |
| $i_{b,min}$ | Lower bound of the battery current (A) |
| $i_c$ | Constant current variable (A) |
| $i_{c,max}$ | Upper bound of the constant current variable (A) |
| $i_{c,min}$ | Lower bound of the constant current variable (A) |
| $I_{ex}$ | Amplitude of the sinusoidal signal (A) |
| $K_{0-4}$ | Coefficients of OCV-SOC curve |
| $m$ | Vehicle mass (kg) |
| $N$ | Time length of the driving cycle |
| $N_{red}$ | Final transmission ratio |
| $P_{bat}$ | Power of the battery (kW) |
| $P_{bat,max}$ | Upper bound of the battery power (kW) |
| $P_{bat,min}$ | Lower bound of the battery power (kW) |
| $P_{dem}$ | Power required to follow the driving cycle (kW) |
| $P_{gen}$ | Power output of the generator (kW) |
| $P_{gen,max}$ | Upper bound of the generator power (kW) |
| $P_{gen,min}$ | Lower bound of the generator power (kW) |
| $P_{mot}$ | Power output of the motor (kW) |
| $Q_b$ | Capacity of battery cell (Ah) |
| $\widehat{Q_b}$ | Estimated capacity of battery cell (Ah) |
| $Q_p$ | Capacity of battery pack (Ah) |
| $R_b$ | Ohmic resistance of battery pack (Ω) |
| $R_s$ | Ohmic resistance of the battery cell (Ω) |
| $\widehat{R_s}$ | Estimated ohmic resistance of the battery cell (Ω) |
| $R_t$ | Resistance of the RC pair (Ω) |
| $\widehat{R_t}$ | Estimated resistance of the RC pair (Ω) |
| $R_{tire}$ | Wheel radius (m) |
| $s$ | Complex Laplace variable |
| $SOC_{max}$ | Upper bound of the recommended SOC usage range |
| $SOC_{min}$ | Lower bound of the recommended SOC usage range |
| $t$ | Time (s) |
| $t_0$ | Initial time (s) |
| $T_s$ | Sampling time (s) |
| $T_p$ | Half period of the injected signal (s) |
| $V$ | Vehicle velocity (m/s) |
| $V_{bf}$ | Filtered terminal voltage of the battery (V) |
| $V_C$ | Voltage across the RC pair (V) |
| $V_{OC}$ | Open-circuit voltage of the battery (V) |
| $V_{nom}$ | Nominal voltage of battery (V) |
| $z$ | Normalized state of charge |
| $z_0$ | Initial battery SOC |
| $\alpha$ | Climbing angle (°) |
| $\gamma$ | Penalty factor |
| $\eta$ | Coulomb efficiency of battery |
| $\eta_{mot}$ | Efficiency of the combination of motor and inverter |
| $\eta_r$ | Regenerative braking system efficiency |
| $\eta_T$ | Transmission efficiency |
| $\rho$ | Air density (kg/m$^3$) |
| $\tau$ | Time constant of the RC pair (s) |
| $\hat{\tau}$ | Estimated time constant of the RC pair (s) |
| $\varphi$ | Instantaneous fuel consumption (g) |
| $\omega_{motor}$ | Motor speed (RPM) |

**Acronyms**

| | |
|---|---|
| DEKF | Dual extended Kalman filter |
| DP | Dynamic programming |
| ECM | Equivalent circuit model |
| EGU | Engine-generator unit |
| EKF | Extended Kalman filter |
| HEV | Hybrid electric vehicle |
| ICE | Internal combustion engine |
| PMS | Power management strategy |
| PMSM | Permanent magnet synchronous machine |
| OCV | Open circuit voltage |
| RMS | Root-Mean-Square |
| SIC | Simultaneous identification and control |
| SOC | State of charge |
| SOH | State of health |

## 1. Introduction

The pursuit to increase the efficiency and reduce the pollution emission of vehicles has called for the development of hybrid electric vehicles (HEVs) [1]. Compared with conventional vehicles, HEVs adopt an extra power source (i.e., batteries) as a buffer to increase overall engine efficiency [2]. For the battery, accurate estimation of its parameters and states including the state of charge (SOC) and the state of health (SOH) is essential for a safe, reliable and efficient operation of HEVs [3]. Condition monitoring of the battery is crucial, as the effective monitoring can help avoid violating the operating constraints (e.g. overcharge/overdischarge) and therefore prolong the battery life [4]. However, parameter estimation and fuel consumption optimization generally conflict because the power management strategy (PMS) minimizing the fuel consumption does not necessarily ensure a battery current profile which contains sufficiently rich information for estimation. Single power source systems (e.g. electric vehicles) can not achieve these two objectives concurrently. However, for the over-actuated system (e.g. HEVs), the extra degree of freedom in power allocation can be used to achieve simultaneous identification and control (SIC) [5]. There exists related work which achieves parameters estimation and output regulation simultaneously in over-actuated systems. For instance, Hasanzadeh et al. focused on induction machines and achieved the identification of rotor resistance and minimization of the torque ripple through the injection of a relatively low-frequency signal [6]. Reed et al. also studied the over-actuated features of permanent magnet synchronous machines in order to simultaneously estimate parameters and regulate torque [7]. Leve and Jah injected permanent excitation for parameter identification without disturbing the control objective after exploiting the "null motion" of over-actuated spacecraft [8]. Similarly, Chen and Wang applied independent motors to inject additional excitation so that the tire-road friction coefficient can be identified accurately given the extra degrees of freedom provided by electric vehicles [9]. This paper proposes SIC for HEV to significantly improve the estimation performance of battery parameters/states with a slight increase of fuel consumption.

Generally, HEVs can be classified into three types according to different configurations of powertrain: 1) series [10]; 2) parallel [10]; 3) power-split [11]. In this paper, the series HEV is studied because the engine is mechanically decoupled from the wheels [12], which means that the engine operating point determined by the engine speed and torque can be optimal. Since HEVs are sophisticated electro-mechanical-chemical systems, a PMS is required. In the existing literature, a large number of PMSs have been designed, which can be divided into two groups: rule-based control and optimization-based control [13]. More and more attention is paid to optimization-based control strategies, which can be classified into two groups [14]: 1) global optimization; 2) real-time optimization. Among most of the PMSs, the variation of battery parameters is neglected which can result in unsafe and inefficient operation of HEVs and motivates us to apply SIC to HEVs.

In order to identify the states and parameters of the battery, the estimation approach, which is based on the equivalent-circuit model (ECM) [15], is applied due to its simplicity and adequate fidelity [16]. SOC estimation approaches include coulomb counting, extended Kalman filter (EKF) [17], unscented Kalman filter [18], $H_\infty$ observer [19], sliding mode observer [20], and fuzzy-logic-based method [21]. SOH is generally defined as the ratio of the remaining capacity to the original capacity [22] and it can also be estimated through similar algorithms, such as EKF [23], least-squares methods [24], moving-horizon observers [25], and Lyapunov-based methods [26]. Since the estimation approaches mentioned above require knowledge of battery parameters, which are often obtained through offline identification, variations in battery parameters due to aging and changes in operating conditions (e.g. temperatures) can cause performance degradation [27]. To address this issue, a dual extended Kalman filter (DEKF) has been developed to estimate SOC, SOH, and battery parameters concurrently [28]. In addition to the estimation algorithms, the input and output signals applied in the estimation process, i.e. battery current and voltage, can dramatically influence the estimation accuracy. Therefore, it's necessary to optimize the battery current waveform in order to ensure signal richness and therefore identification accuracy [29], [30].

Given the extra degree of freedom in PMSs offered by HEVs, it is possible to inject sufficiently rich signals and achieve desired powertrain torques simultaneously. In order to optimize SIC performance, an innovative DP-based method is adopted in this paper. Since the sequential algorithm, which uses frequency-scale separation and estimates parameters/states sequentially through active current injection, is proved to be more effective than DEKF, which concurrently identify parameters and states, according to the experimental results [31], it is integrated in the DP framework to improve estimation performance. To the best of our knowledge, this is the first paper exploiting the over-actuated feature of HEV to inject active signals into the battery for better estimation performance. Even though the signal injection induces a slight increase in fuel consumption, accurate identification of battery states and parameters is significant for the safe, efficient, and reliable operation of Lithium-ion batteries [32].

The rest of the paper is organized as follows. In Section 2, the model of the series HEV is presented. The novel DP is developed and presented in Section 3 along with a review of the sequential algorithm. In Section 4, simulation results of the proposed SIC and the comparison with the baseline results without active signal injection are discussed. Conclusions are given in Section 5.

## 2. Modeling of Series Hybrid Electric Vehicle

The series HEV, whose architecture is shown in Fig. 1 [33], is studied in this paper just as an example and the similar technology of SIC can work on all topologies. The series powertrain is composed of an internal combustion engine (ICE), a generator, a battery, a motor, and an associated inverter. The ICE together with the generator is called the engine-generator unit (EGU). The motor is mechanically coupled to the wheels. Since there are two power sources (i.e. EGU and battery), which provide the over-actuated feature, the model satisfies the following power balance equations:

$$\begin{aligned} P_{dem}(t) &= P_{mot}(t) \\ P_{mot}(t) &= \eta_{mot}(t)\left[P_{gen}(t) + P_{bat}(t)\right] \end{aligned} \quad (1)$$

where $P_{dem}(t)$, detailed below in (2)-(4), is the power required to follow the driving cycle, $P_{mot}(t)$ is the power output of the motor, $\eta_{mot}(t)$, illustrated in Fig.2 [34], is the efficiency of the combination of motor and inverter, which is a function of motor speed and torque, $P_{gen}(t)$ is the output power of the generator, and $P_{bat}(t)$ is the battery power.

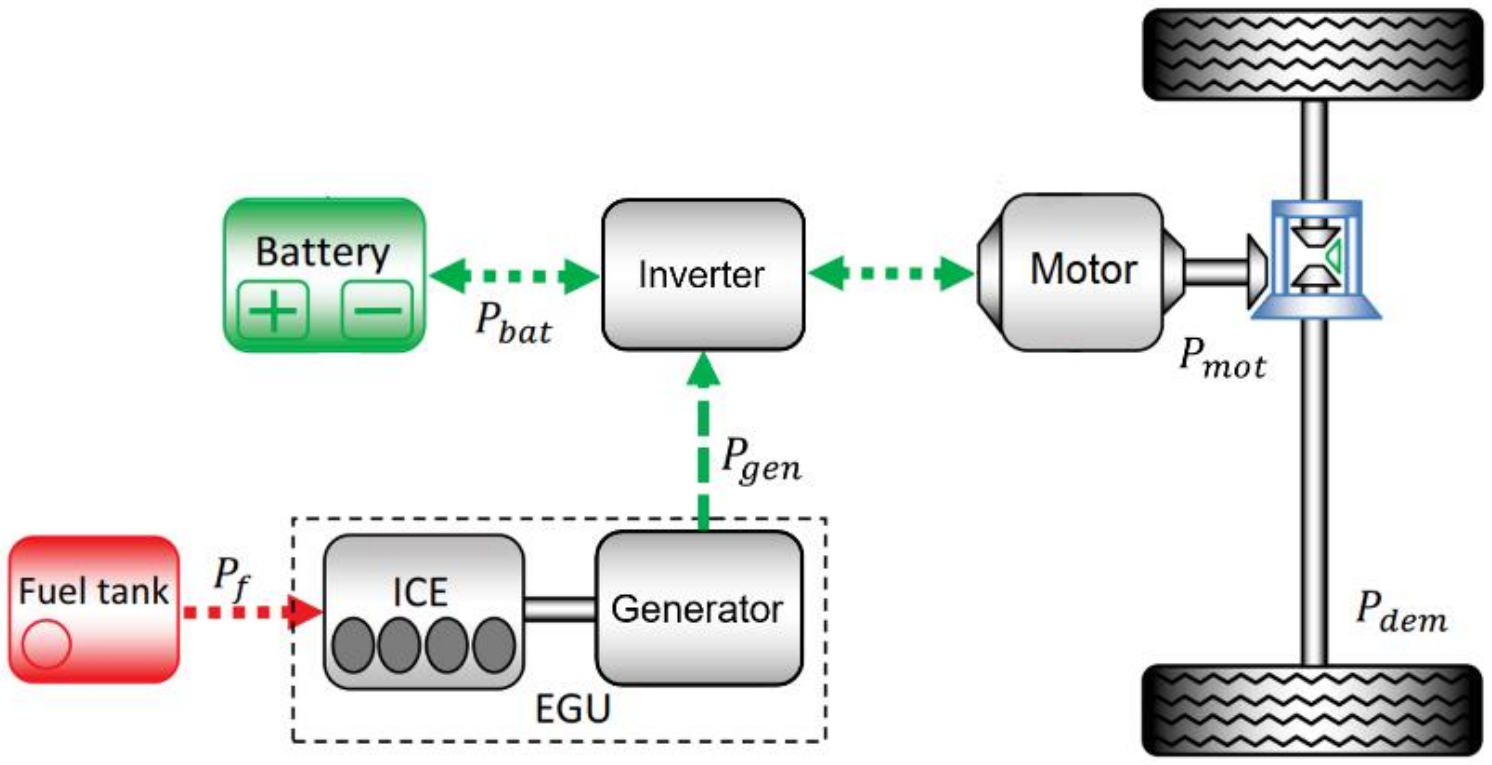

Fig. 1. The schematic of the series HEV powertrain.

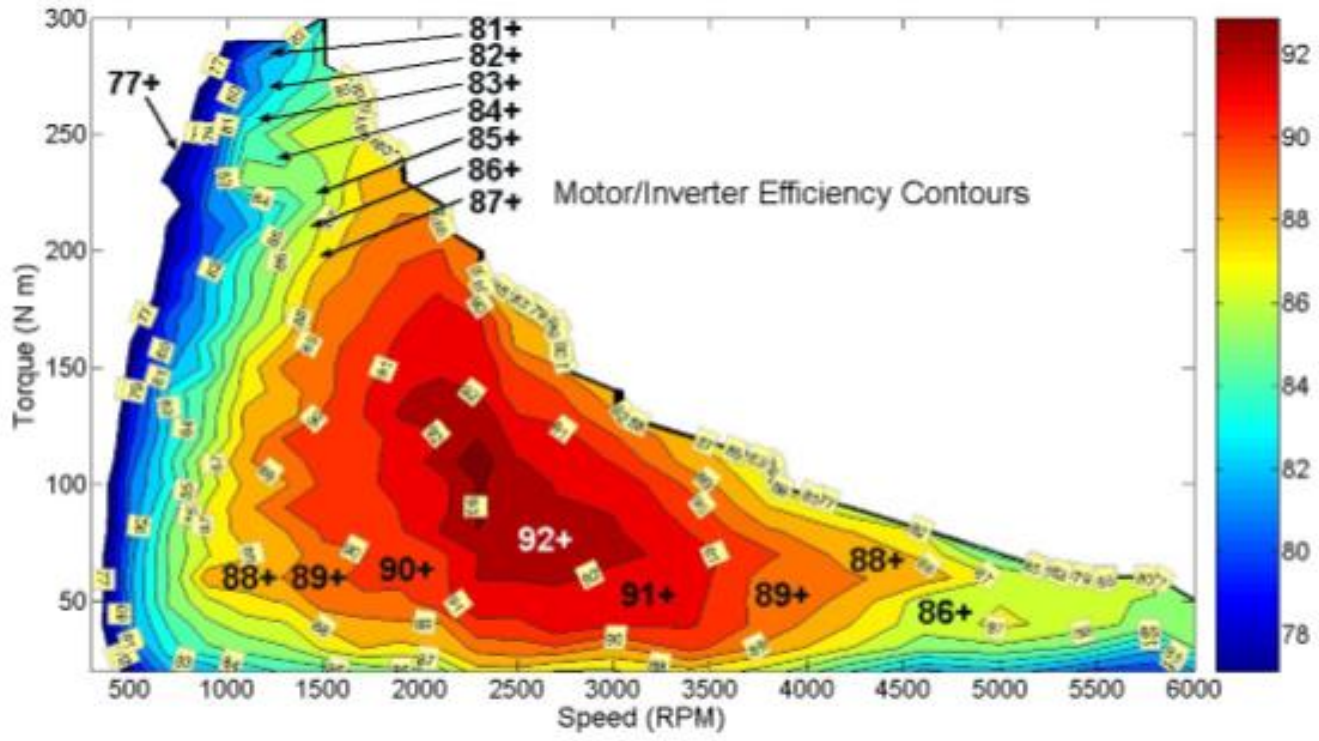

Fig. 2. Motor/Inverter efficiency contours.

### 2.1 Vehicle Model

The main parameters of the studied vehicle under simulation are summarized in Table 1 [35] and the basic dynamic model is given as:

$$V = \omega_{motor} \times N_{red} \times R_{tire} \tag{2}$$

(1) Traction Mode:

$$mgfV \cos\alpha + 0.5 C_D \rho A_{front} V^3 + mV\dot{V} + mgV \sin\alpha = P_{dem}\eta_T \tag{3}$$

(2) Regenerative Braking Mode:

$$mgfV \cos\alpha + 0.5 C_D \rho A_{front} V^3 + mV\dot{V} + mgV \sin\alpha = P_{dem}/\eta_r \tag{4}$$

where $V$ is the vehicle velocity, $\omega_{motor}$ is the motor speed, $N_{red}$ is the final transmission ratio, $R_{tire}$ is the wheel radius, $m$ is the vehicle mass, $g$ is the gravitational acceleration, $f$ is the rolling resistance coefficient, $\alpha$ is the climbing angle, $C_D$ is the air drag coefficient, $\rho$ is the air density, $A_{front}$ is the frontal area, $\eta_T$ is the transmission efficiency, and $\eta_r$ is the regenerative braking system efficiency.

Table 1. Basic Parameters of the Series HEV.

| Parameters | Values (Unit) |
|---|---|
| Mass of Vehicle ($m$) | 1254 kg |
| Wheel Radius ($R_{tire}$) | 0.287 m |
| Frontal Area ($A_{front}$) | 2.52 m² |
| Air Drag Coefficient ($C_D$) | 0.3 |
| Rolling Resistance Coefficient ($f$) | 0.015 |
| Transmission Efficiency ($\eta_T$) | 0.9 |
| Regenerative Braking System Efficiency ($\eta_r$) | 0.25 |
| Final Transmission Ratio ($N_{red}$) | 4.113 |
| Nominal Voltage of Battery ($V_{nom}$) | 201.6V |
| Capacity of Battery Pack ($Q_p$) | 6.5Ah |
| Ohmic Resistance ($R_b$) | ~0.5Ω |

## 2.2 Battery Model

The main parameters of the battery pack are listed in Table 1. The battery behavior in the PMS is represented by the Rint model for simplicity [35], as the ohmic resistance of battery will dominate the power loss. As for the estimation problem, the first-order ECM is adopted in this paper, which is shown in Fig. 3.

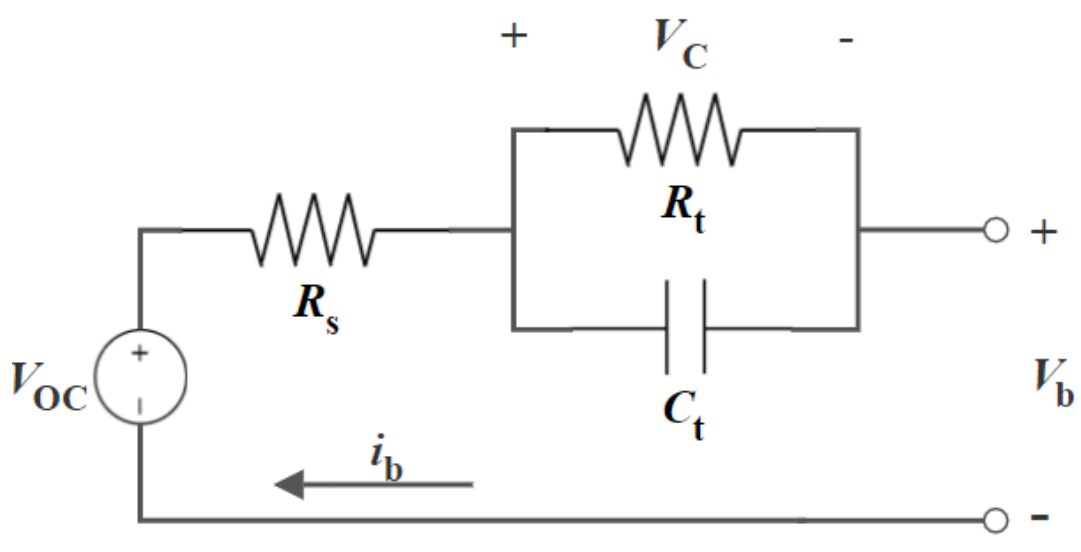

Fig. 3. First-order equivalent circuit model.

The battery terminal voltage is represented by $V_b$ and the battery current is defined as $i_b$ (positive for discharging and negative for charging). Although higher-order models (e.g., 2RC and 3RC circuit models as well as the electrochemical model) can represent the battery dynamics more accurately than the first-order ECM, there is a trade-off between model accuracy and computational cost [36]. The parameter estimation of higher-order models has a much higher computational cost and requires more richness of the battery current waveform. For instance, if a second-order ECM is adopted, there are 7 parameters to be identified; consequently, at least four sinusoidal current components are required to fulfill the richness condition [37], which is difficult to realize in practical applications. Therefore, the first-order model is adopted in this paper considering both model accuracy and simplicity. According to Kirchhoff's law, the ECM has the following dynamics:

$$\begin{cases} \dot{V}_C = -\frac{1}{C_t R_t} V_C + \frac{1}{C_t} i_b \\ V_b = V_{OC} - R_s i_b - V_C \end{cases}, \tag{5}$$

where $V_C$ is the voltage across the RC pair, $C_t$ and $R_t$ are the capacitance and resistance of the RC pair, $R_s$ is the ohmic resistance of the battery cell, $V_{OC}$ is the open-circuit voltage (OCV) of the battery. The OCV-SOC relationship can be represented by then following equation [38]:

$$V_{OC}(z) = K_0 - \frac{K_1}{z} - K_2 z + K_3 \ln(z) + K_4 \ln(1 - z), \quad (6)$$

where $K_{0-4}$ are the model parameters, and $z$ is the normalized SOC (from 0 to 1), which can be calculated using the following model:

$$z = z_0 - \int_{t_0}^{t} \frac{\eta}{Q_b} i_b(\tau)\, d\tau, \quad (7)$$

where $z_0$ is the initial SOC, $\eta$ is the coulomb efficiency, $t_0$ is the initial time, $t$ is the time, and $Q_b$ is the battery cell capacity. Since lithium-ion batteries have good energy density and long cycling life, the Samsung 18650 Lithium battery cell is applied and studied. Static capacity Hybrid pulse tests are conducted at 20°C to determine the parameters [36]. The model parameters $K_{0-4}$ are determined to be 2.6995, 0.0574, -1.3967, -0.55018, -0.0377 respectively [38]. The other specifications and parameters of the chosen battery cell are listed in Table 2.

Table 2. Basic parameters of the Samsung 18650 Lithium battery cell.

| **Parameters** | **Value (Unit)** |
|---|---|
| Nominal Voltage | 3.63V |
| Cell Capacity ($Q_b$) | 2.47Ah |
| Cell Weight | 45g |
| Ohmic Resistance ($R_s$) | ~100mΩ |
| Diffusion Resistance ($R_t$) | ~30mΩ |
| Time Constant of RC Pair(τ) | ~15s |
| Coulomb Efficiency (η) | 0.98 |
| Standard Deviation of Voltage Measurement Noise (σV) | 20mV |

## 3. Review of the Sequential Algorithm

The estimation of battery states (e.g. SOC and SOH) and parameters is an important, though difficult, task. According to Cramer-Rao Bound analysis, the sequential algorithm is more accurate than estimation algorithms which identify all the battery parameters and states concurrently [31]. It is because the sequential algorithm divides the estimation process into several steps, which can reduce the uncertainties introduced during the identification [31]. In order to simplify the calculations, Eq. (6) is linearized because the slope of the OCV-SOC curve is constant within the normal operating range [39]:

$$V_{OC}(z) = a\left(z_0 - \int_{t_0}^{t} \frac{\eta}{Q_b} i_b(\tau)\, d\tau\right) + b, \tag{7}$$

where $a$ and $b$ are the coefficients of the linearized OCV-SOC curve. After applying the Laplace Transform, the battery dynamics in the frequency domain are obtained as shown below:

$$V_b(s) = \frac{az_0+b}{s} - \frac{a}{s}\frac{\eta}{Q_b} i_b(s) - R_s i_b(s) - \frac{R_t}{1+\tau s} i_b(s), \tag{8}$$

where $s$ is the complex Laplace variable. The Laplace Transform of the battery terminal voltage is therefore composed of four parts, which are determined by the initial SOC, capacity, ohmic resistance, and RC pair respectively. Therefore, it's possible to separate the four components through filtering. The first term, related to the initial SOC, can be removed by a high-pass filter because it is constant in the time domain. After filtering, the terminal voltage, i.e. $V_{bf}$, is dominated by the ohmic resistance when the current frequency is "high". Therefore, the sequential algorithm can be summarized in three steps as shown in Fig. 4 [31]:

Step #1: Based on a high-pass filter and the injected high-frequency current, the ohmic resistance is estimated using the extended Kalman filter (EKF).

Step #2: Applying the identified ohmic resistance, the parameters of the RC pair (e.g. the resistance $R_t$ and the capacitance $C_t$) are estimated using EKF when the high-pass filter is incorporated and the medium-frequency current is injected.

Step #3: Adopting the identified $R_s$, $R_t$, and $C_t$, the SOC and SOH are simultaneously estimated based on the unfiltered system using the DEKF.

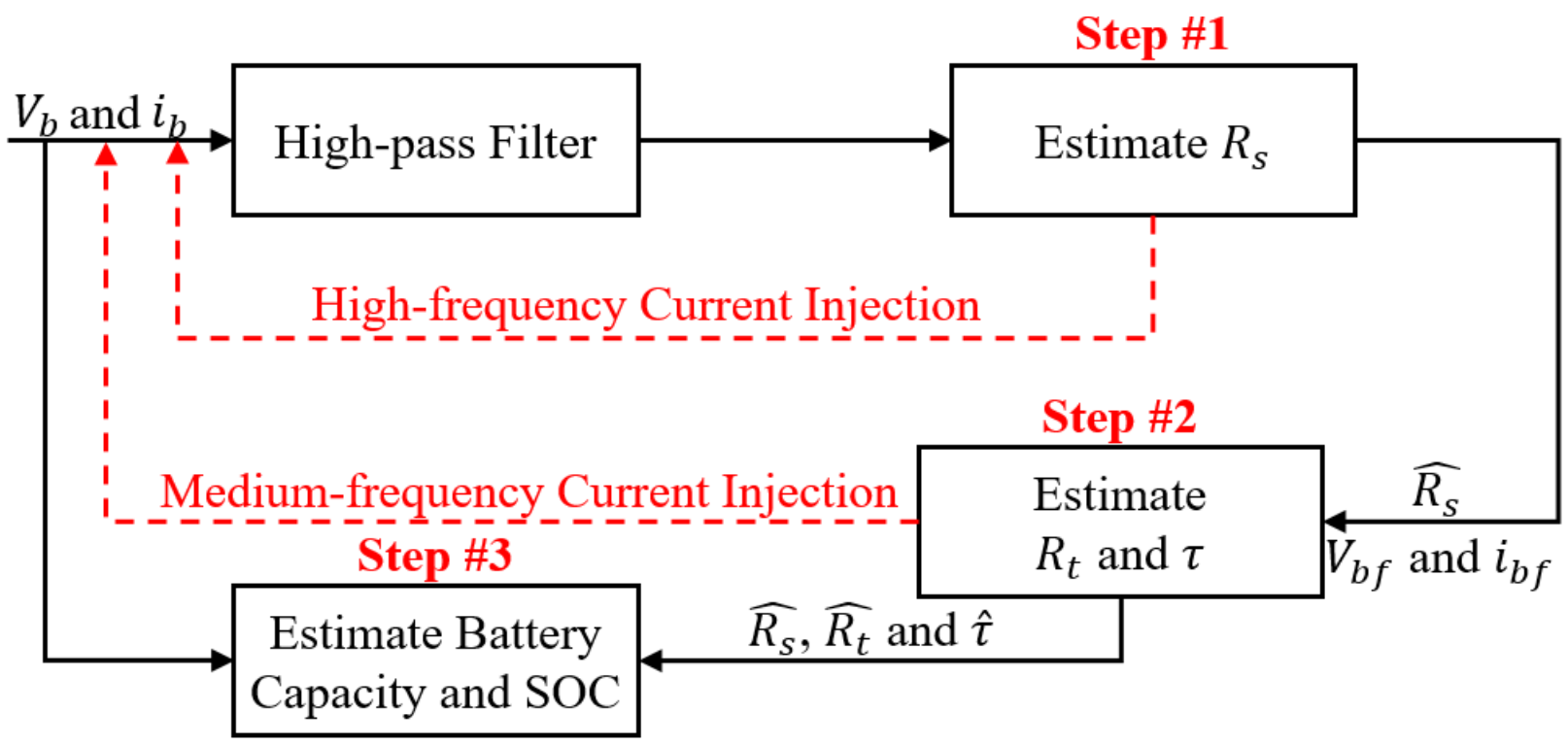

Fig. 4. Flow chart of sequential algorithm.

The following is a brief review of EKF and DEKF. The EKF, which is one of the most common estimation algorithms, is used in both Step #1 and Step #2 of the sequential algorithm in order to avoid the effect of process and measurement noise [40]. The general state-space equation of the discrete-time EKF can be expressed as follows:

$$\begin{cases} \boldsymbol{\theta}_{k+1} = \boldsymbol{\theta}_k + \boldsymbol{r}_k \\ \mathbf{X}_{k+1} = \mathbf{H}(\mathbf{X}_k, \boldsymbol{\theta}_k, \mathbf{u}_k) + \boldsymbol{w}_k, \\ \mathbf{Y}_{k+1} = \mathbf{G}(\mathbf{X}_k, \boldsymbol{\theta}_k, \mathbf{u}_k) + \boldsymbol{v}_k \end{cases} \tag{9}$$

where $k$ is the time index, $\boldsymbol{\theta}_k$ is the parameter vector, $\boldsymbol{r}_k$ is the process noise for parameters, $\mathbf{X}_k$ is the state vector, $\mathbf{u}_k$ is the input vector, $\boldsymbol{w}_k$ is the process noise for states, $\mathbf{Y}_k$ is the output vector, and $\boldsymbol{v}_k$ is the measurement noise for outputs. Since the estimation of states is not involved in Step #1-2, the calculation process is shown below:

1) Initialization:

$$\begin{cases} \widehat{\boldsymbol{\theta}}_0 = E[\boldsymbol{\theta}_0] \\ \Sigma_{\boldsymbol{\theta}_0} = E\left[\left(\boldsymbol{\theta}_0 - \widehat{\boldsymbol{\theta}}_0\right)\left(\boldsymbol{\theta}_0 - \widehat{\boldsymbol{\theta}}_0\right)^T\right], \end{cases} \tag{10}$$

where $\sum_{\boldsymbol{\theta}_0}$ is the covariance matrix of parameter estimation error.

2) Parameter prediction:

$$\begin{cases} \widehat{\boldsymbol{\theta}}_k^- = \widehat{\boldsymbol{\theta}}_{k-1} \\ \Sigma_{\boldsymbol{\theta}_k}^- = \Sigma_{\boldsymbol{\theta}_{k-1}} + \Sigma_{r_{k-1}} \end{cases}, \tag{11}$$

where $\sum_{r_{k-1}}$ is the covariance matrix of process noise.

3) Parameter update:

$$\begin{cases} \mathbf{K}_k^{\boldsymbol{\theta}} = \sum_{\boldsymbol{\theta}_{k-1}}^{-} \left(\mathbf{C}_{k-1}^{\boldsymbol{\theta}}\right)^T \left[\left(\mathbf{C}_{k-1}^{\boldsymbol{\theta}}\right) \sum_{\boldsymbol{\theta}_{k-1}}^{-} \left(\mathbf{C}_{k-1}^{\boldsymbol{\theta}}\right)^T + \sum_{r_{k-1}}\right] \\ \hat{\boldsymbol{\theta}}_k = \hat{\boldsymbol{\theta}}_k^{-} + \mathbf{K}_k^{\boldsymbol{\theta}} \left[\mathbf{Y}_k - \mathbf{G}\left(\mathbf{X}_{k-1}, \hat{\boldsymbol{\theta}}_k^{-}, \mathbf{u}_k\right)\right] \\ \sum_{\boldsymbol{\theta}_k} = \left(\mathbf{I} - \mathbf{K}_k^{\boldsymbol{\theta}} \mathbf{C}_{k-1}^{\boldsymbol{\theta}}\right) \sum_{\boldsymbol{\theta}_{k-1}}^{-} \end{cases}, \tag{12}$$

where $\mathbf{C}_{k-1}^{\boldsymbol{\theta}} = \frac{\partial \mathbf{G}(\mathbf{X}_{k-1}, \boldsymbol{\theta}, \mathbf{u}_k)}{\partial \boldsymbol{\theta}}\Big|_{\boldsymbol{\theta} = \hat{\boldsymbol{\theta}}_k^{-}}$.

In Step #3, since both the SOC and SOH of the battery is estimated, the DEKF is applied, which is a commonly used technique to estimates states and parameters concurrently [41]. The state-space equation is the same as that of EKF, as shown in Eq. (9). The detailed calculation process is summarized as follows:

1) Initialization:

$$\begin{cases} \hat{\boldsymbol{\theta}}_0 = E[\boldsymbol{\theta}_0] \\ \sum_{\boldsymbol{\theta}_0} = E\left[\left(\boldsymbol{\theta}_0 - \hat{\boldsymbol{\theta}}_0\right)\left(\boldsymbol{\theta}_0 - \hat{\boldsymbol{\theta}}_0\right)^T\right] \\ \hat{\mathbf{X}}_0 = E[\mathbf{X}_0] \\ \sum_{\mathbf{X}_0} = E\left[\left(\mathbf{X}_0 - \hat{\mathbf{X}}_0\right)\left(\mathbf{X}_0 - \hat{\mathbf{X}}_0\right)^T\right] \end{cases}, \tag{13}$$

where $\sum_{\boldsymbol{\theta}_0}$ is the covariance matrix of parameter estimation error and $\sum_{\mathbf{X}_0}$ is the covariance matrix of state estimation error.

2) Parameter prediction:

$$\begin{cases} \hat{\boldsymbol{\theta}}_k^{-} = \hat{\boldsymbol{\theta}}_{k-1} \\ \sum_{\boldsymbol{\theta}_k}^{-} = \sum_{\boldsymbol{\theta}_{k-1}} + \sum_{r_{k-1}} \end{cases}, \tag{14}$$

where $\sum_{r_{k-1}}$ is the covariance matrix of process noise.

3) State prediction:

$$\begin{cases} \hat{\mathbf{X}}_k^{-} = \mathbf{H}\left(\hat{\mathbf{X}}_{k-1}, \hat{\boldsymbol{\theta}}_k^{-}, \mathbf{u}_k\right) \\ \sum_{\mathbf{X}_k}^{-} = \mathbf{A}_k \sum_{\mathbf{X}_{k-1}} \mathbf{A}_k^T + \sum_{w_k} \end{cases}, \tag{15}$$

where $\mathbf{A}_k = \frac{\partial \mathbf{H}(\mathbf{X}, \hat{\boldsymbol{\theta}}_k^{-}, \mathbf{u}_k)}{\partial \mathbf{X}}\Big|_{\mathbf{X} = \hat{\mathbf{X}}_k^{-}}$.

4) State update:

$$\begin{cases} \mathbf{K}_k^{\mathbf{X}} = \sum_{\mathbf{X}_k}^{-} \left(\mathbf{C}_k^{\mathbf{X}}\right)^T \left[\left(\mathbf{C}_k^{\mathbf{X}}\right) \sum_{\mathbf{X}_k}^{-} \left(\mathbf{C}_k^{\mathbf{X}}\right)^T + \sum_{r_k}\right] \\ \hat{\mathbf{X}}_k = \hat{\mathbf{X}}_k^{-} + \mathbf{K}_k^{\mathbf{X}} \left[\mathbf{Y}_k - \mathbf{G}\left(\hat{\mathbf{X}}_k^{-}, \hat{\boldsymbol{\theta}}_k^{-}, \mathbf{u}_k\right)\right] \\ \sum_{\mathbf{X}_k} = \left(\mathbf{I} - \mathbf{K}_k^{\mathbf{X}} \mathbf{C}_k^{\mathbf{X}}\right) \sum_{\mathbf{X}_k}^{-} \end{cases}, \tag{16}$$

where $\mathbf{C}_k^{\mathbf{X}} = \frac{\partial \mathbf{G}(\mathbf{X}, \hat{\boldsymbol{\theta}}_k^{-}, \mathbf{u}_k)}{\partial \mathbf{X}}\Big|_{\mathbf{X} = \hat{\mathbf{X}}_k^{-}}$.

5) Parameter update:

$$\begin{cases} \mathbf{K}_k^{\theta} = \Sigma_{\theta_k}^{-}\left(\mathbf{C}_k^{\theta}\right)^T\left[\left(\mathbf{C}_k^{\theta}\right)\Sigma_{\theta_k}^{-}\left(\mathbf{C}_k^{\theta}\right)^T + \Sigma_{r_k}\right] \\ \hat{\theta}_k = \hat{\theta}_k^{-} + \mathbf{K}_k^{\theta}\left[\mathbf{Y}_k - \mathbf{G}\left(\hat{\mathbb{X}}_k, \hat{\theta}_k^{-}, \mathbf{u}_k\right)\right] \\ \Sigma_{\theta_k} = \left(\mathbf{I} - \mathbf{K}_k^{\theta}\mathbf{C}_k^{\theta}\right)\Sigma_{\theta_k}^{-} \end{cases}, \tag{17}$$

where $\mathbf{C}_k^{\theta} = \frac{\mathrm{d}\mathbf{G}(\hat{\mathbb{X}}_k, \theta, \mathbf{u}_k)}{\mathrm{d}\theta}\Big|_{\theta=\hat{\theta}_k^{-}}$.

## 4. Simultaneous Estimation and Optimization using Dynamic Programming

The DP approach is adopted in this paper to balance the estimation and optimization objectives as they are generally conflicting with each other. The baseline DP without active signal injection, denoted as DP-, is discussed firstly as follows:

$$J_{DP-} = \sum_{k=1}^{N} \varphi\left(P_{gen}(k) \times T_s\right) + \gamma \Delta_{SOC}, \tag{18}$$

and

$$\Delta_{SOC} = \begin{cases} 0 & , SOC(N) \geq SOC(1) \\ SOC(N) - SOC(1) & , SOC(N) < SOC(1) \end{cases}$$

subject to the constraints:

$$SOC_{min} \leq SOC(k) \leq SOC_{max}$$

$$SOC(N) = SOC(1)$$

$$P_{bat,min} \leq P_{bat}(k) \leq P_{bat,max}$$

$$P_{gen,min} \leq P_{gen}(k) \leq P_{gen,max}$$

$$P_{gen}(k) + P_{bat}(k) = P_{dem}(k)/\eta_{mot}(k)$$

where $\varphi$ is the instantaneous fuel consumption, which is illustrated in Fig. 5 [42], $k$ is the time index, $T_s$ is the sampling time, $N$ is the time length of the driving cycle, $\gamma$ is the penalty factor to force the final SOC to equal to the initial value, $SOC_{min}$ and $SOC_{max}$ are the lower and upper bounds of the recommended SOC usage range, i.e. 0.2-0.9, $P_{bat,min}$ and $P_{bat,max}$ are the lower and upper bounds of the battery power, and $P_{gen,min}$ and $P_{gen,max}$ are the lower and upper bounds of the generator power, repectively. Since the engine is mechanically decoupled from the wheels for series HEVs, it is reasonable to assume that the working point of the engine, which is defined by the engine torque and speed, always lies on the basic operating line [42], as shown in Fig. 5, in order to maximize efficiency. After acquiring the current signal of DP-, the parameters and states of the battery are concurrently estimated using DEKF for comparison.

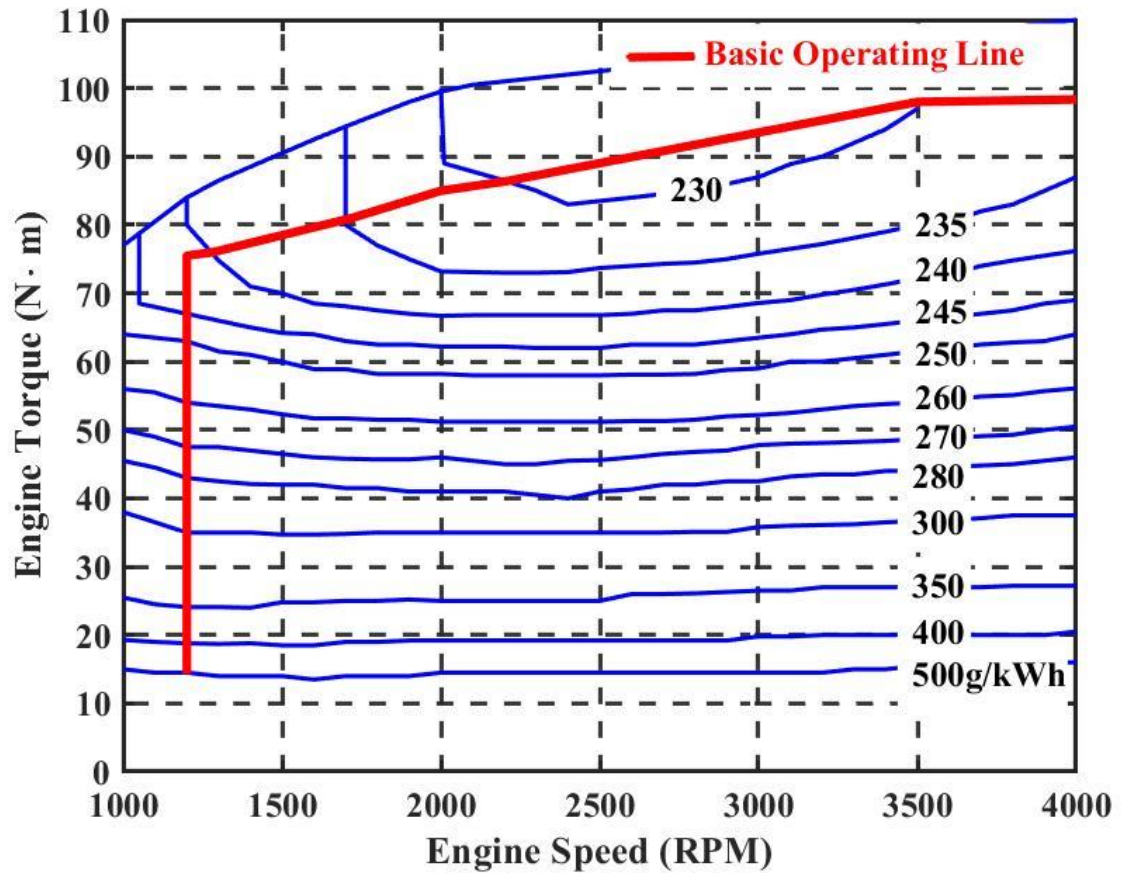

Fig. 5. Fuel consumption contours.

When the current signal is injected, the constraints of DP need to be changed. Since a sinusoidal battery current is injected, the formula of the total battery current can be represented by Eq. (19):

$$i_b(k) = I_{ex}\, cos\left(2\pi f_{injected} T_s k\right) + i_c(j), \tag{19}$$

where $I_{ex}$ is the amplitude of the sinusoidal signal, $f_{injected}$ is the frequency of the injected signal, which is chosen to be 0.5Hz and 0.05Hz, $i_c$ is a variable which changes with $j$, a time index with different time interval. According to the simulation results, the existence of $i_c(j)$ can prevent the sinusoidal signal, i.e. $I_{ex}\, cos\left(2\pi f_{injected} T_s k\right)$, from being compensated during the process of DP. On the other hand, $i_c$ will affect the control of HEVs negatively if the time interval of $j$ is much longer than $T_s$. The time interval for $i_c$ to remain constant was finally chosen to be one half of the period of the sinusoidal signal. The battery power $P_{bat}(k)$ can be formulated as Eq. 20 because when considering the PMS for HEVs, the Rint battery model can be used [35],

$$P_{bat}(k) = V_{nom} i_b(k) - {i_b}^2(k) R_b. \tag{20}$$

The novel cost function $J_{DP+}$ for the DP with active signal injection, denoted as DP+, is then formulated as

$$J_{DP+} = \sum_{k=1}^{N} \varphi\left(P_{gen}(k) \times T_s\right) + \gamma[SOC(N) - SOC(1)], \tag{21}$$

subject to the constraints:

$$SOC_{min} \leq SOC(k) \leq SOC_{max}$$

$$SOC(N) = SOC(1)$$

$$P_{gen,min} \le P_{gen}(k) \le P_{gen,max}$$

$$P_{bat,min} \le P_{bat}(k) \le P_{bat,max}$$

$$P_{gen}(k) + P_{bat}(k) = P_{dem}(k)/\eta_{mot}(k)$$

$$i_{c,min}(j) \le i_c(j) \le i_{c,max}(j)$$

where $i_{c,min}(j)$ and $i_{c,max}(j)$, defined by Eq. (22), are the lower and upper bounds of the variable $i_c$.

$$\begin{cases} i_{c,min}(j) = \max\limits_{T_p(j-1)\le T_s k \le T_p j}\left[i_{b,min} - I_{ex}\cos\left(2\pi f_{injected} T_s k\right)\right] \\ i_{c,max}(j) = \min\limits_{T_p(j-1)\le T_s k \le T_p j}\left[i_{b,max} - I_{ex}\cos\left(2\pi f_{injected} T_s k\right)\right] \end{cases}, \tag{22}$$

$T_p$ is the half period of the injected signal, and $i_{b,min}$ and $i_{b,max}$ are the lower and upper bounds of the total battery current, respectively. The sampling time $T_s$ is set to be 0.2s and 1s for injected signals of 0.5Hz and 0.05Hz, respectively.

With the new battery current signal obtained from DP+, the sequential algorithm discussed in Section 3 is applied and the detailed formulas are presented below:

Step #1: Due to the application of the high-pass filter and high-frequency current signal, the terminal voltage based on Eq. (8) can be simplified as

$$V_{bf}(s) = -R_s i_{bf}(s). \tag{23}$$

where $V_{bf}$ and $i_{bf}$ are the filtered terminal voltage and current of the battery, $R_s$ is the ohmic resistance of the battery cell. In order to estimate the ohmic resistance, the state-space equation of EKF can be formulated as

$$\begin{cases} \widehat{R_s}(k) = \widehat{R_s}(k-1) + r_k \\ V_{bf}(k) = -\widehat{R_s} i_{bf}(k) + v_k \end{cases}. \tag{24}$$

Step #2: After injecting the medium-frequency signal, Eq. (8) can be transformed into Eq. (25) because the ohmic resistance and RC pair dynamics will govern the terminal voltage.

$$V_{bf}(s) = -R_s i_{bf}(s) - \frac{R_t}{1+\tau s} i_{bf}(s). \tag{25}$$

Adopting the estimated value of the ohmic resistance in Step #1, the state-space equation is shown as follows:

$$\begin{cases} \boldsymbol{\theta}_2(k) = \boldsymbol{\theta}_2(k-1) + \boldsymbol{r}_k \\ V_{bf}(k) = -\widehat{R_s}(k) i_{bf}(k) - \widehat{R_t} i_2(k) + \mathbf{v}_k \end{cases}, \tag{26}$$

where

$$\begin{cases} \boldsymbol{\theta}_2(k) = \left[\widehat{R_t}(k) \quad \hat{\tau}(k)\right]^T \\ i_2(k) = \frac{T_s}{T_s+2\hat{\tau}}\left[i_{bf}(k) + i_{bf}(k-1)\right] - \frac{T_s-2\hat{\tau}}{T_s+2\hat{\tau}} i_2(k-1) \end{cases}.$$

Step #3: Applying the ohmic resistance and parameters of the RC pair, the state-space equation for the SOC/SOH co-estimation can be presented below:

$$\begin{cases} \widehat{Q_b}(k) = \widehat{Q_b}(k-1) + r_k \\ \mathbf{X}_3(k) = \begin{bmatrix} e^{-\frac{T_s}{\hat{\tau}}} & 0 \\ 0 & 1 \end{bmatrix} \mathbf{X}_3(k-1) + \begin{bmatrix} \widehat{R_t}\left(1 - e^{-\frac{T_s}{\hat{\tau}}}\right) \\ -\frac{\eta T_s}{\widehat{Q_b}} \end{bmatrix} i_b(k), \\ V_b(k) = V_{OC}\big(z(k)\big) - V_C(k) - \widehat{R_s}(k) i_b(k) \end{cases} \quad (27)$$

where

$$\mathbf{X}_3(k) = [V_C(k) \quad z(k)]^T.$$

The voltage of the RC pair (i.e. $V_C$) is also estimated in the sequential algorithm to improve the estimation accuracy [43].

## 5. Simulation results

Simulations of the series HEV model are conducted using the Urban Dynamometer Driving Schedule (UDDS). As shown in Fig. 6, according to the speed profile of UDDS and Eq. (2)-(4), the power demand profile for the series HEV is obtained. Some basic information about UDDS is listed as follows: 1) the cycle time is 1370s; 2) the maximum vehicle speed is 56.7 MPH; 3) the driving distance is 7.45 miles.

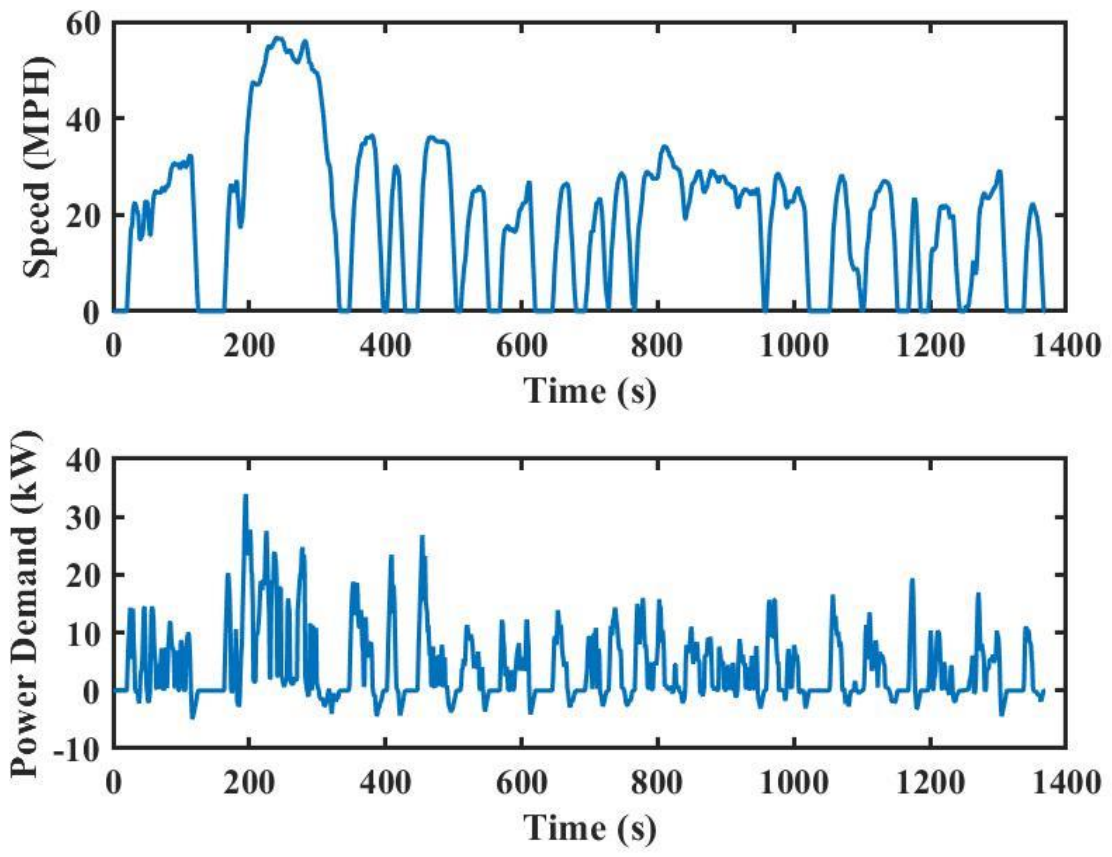

Fig. 6. Speed and power demand profile for the series HEV along UDDS.

## 5.1 Tradeoff between parameters estimation and system optimization

Since a larger signal amplitude can improve estimation accuracy but increase fuel consumption, there is a tradeoff between different objectives. In order to investigate this relationship, the amplitudes of the injected current signal (i.e. $I_{ex}$) are set to range from 0.5A to 10A for the comparison. Since the driving distance for one driving cycle is only 7.45 miles, the simulations are conducted over five consecutive cycles. According to experimental results [31], the effectiveness of the sequential algorithm is verified and it takes less than 200s for the estimated ohmic resistance to converge. Therefore, the 0.5Hz active signal is only injected for the first 200s of the process to minimize the increase of fuel consumption, and the ohmic resistance is identified based on the simulation results. The Root-Mean-Square (RMS) of the estimation error is calculated to indicate the identification accuracy. As shown in Fig. 7(a), the estimation accuracy can be improved by increasing $I_{ex}$ but the fuel consumtion increases as a cost. However, the benefit of increasing $I_{ex}$ is not significant when $I_{ex}$ is beyond a transition area. The “knee point” can be defined when $I_{ex}$ = 6A, and this specific amplitude is chosen for the active signal injection in the following section. The estimation results when injecting signals with different amplitudes are also compared, as shown in Fig. 7(b). This reveals that the signal with a larger amplitude can provide richer information for parameters/states estimation given that the estimated parameter converge to the actual value faster.

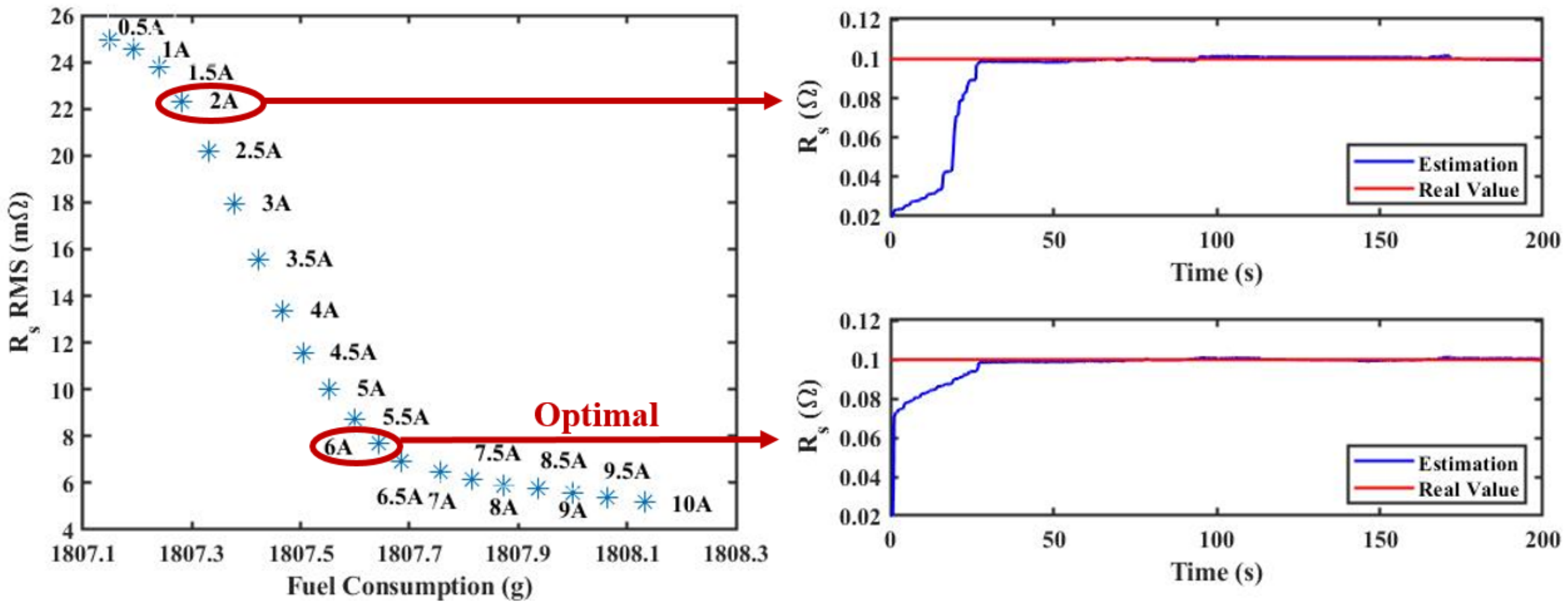

(a) (b)

Fig. 7. Tradeoff between estimation accuracy and fuel economy.

## 5.2 System Optimization

According to the power demand profile shown in Fig. 6, the baseline DP- is applied to solve the energy management problem. The penalty factor $\gamma$ is set to be 350 in the simulation. As shown in Fig. 8(a), the total fuel consumption of the baseline is 1795.13g when no signal is injected. The SOC of the battery fluctuates in a narrow range because the HEV is not plug-in and the battery capacity is small.

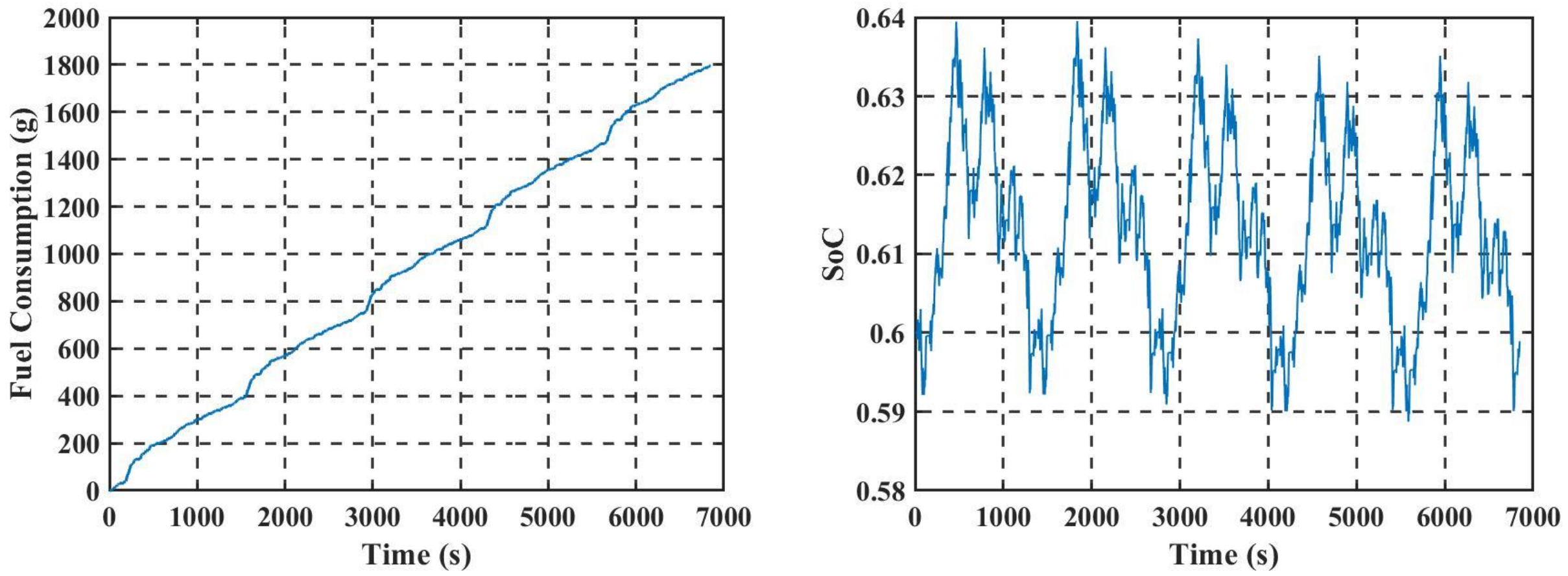

(a) Fuel consumption of engine. (b) SOC of battery.

Fig. 8. Simulation results of DP- without active signal injection.

Then, a sinusoidal signal with the frequency of 0.5Hz is injected for the first 200s. According to Fig. 9(a), the total fuel consumption is 1807.64g, increasing by 0.69% when compared to the one of baseline DP. The SOC profile shows that the battery keeps being charged for the first 200s, showing that the engine supplies power to both the vehicle for following the driving cycle and the battery for actively injecting the signal. Compared with the simulation results of DP-, where the battery starts to provide power to the vehicle at the very beginning, the results of DP+ show that the battery needs to be charged initially to ensure the signal can be injected when the power demand is 0. Specifically, the engine is turned on during idle periods (i.e. zero power demand) to provide the excitation signals with enough richness to the battery for the identification purpose. In addition, the SOC fluctuates in a wide range, which means that the battery is effectively used.

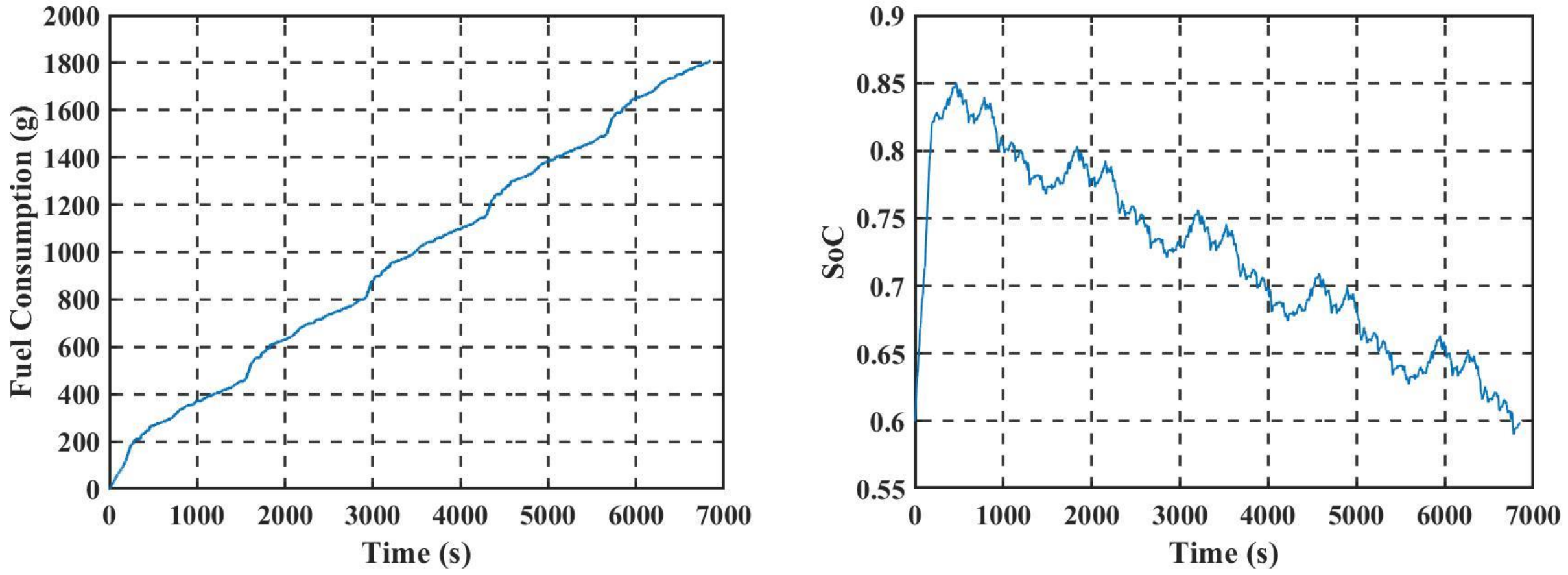

(a) Fuel consumption of engine. (b) SOC of battery.

Fig. 9. Simulation results of DP+ with 0.5Hz Active Signal Injection.

Similarly, a sinusoidal signal with a frequency of 0.05Hz is injected. The injection period is set to be 500s according to the experimental results [31]. The simulation results shown in Fig. 10(a) illustrates that the total fuel consumption for five driving cycles is 1825.02g, increasing by 1.67% when compared with the baseline DP. As shown in Fig. 10(b), the SOC of the battery also fluctuates over a wide range.

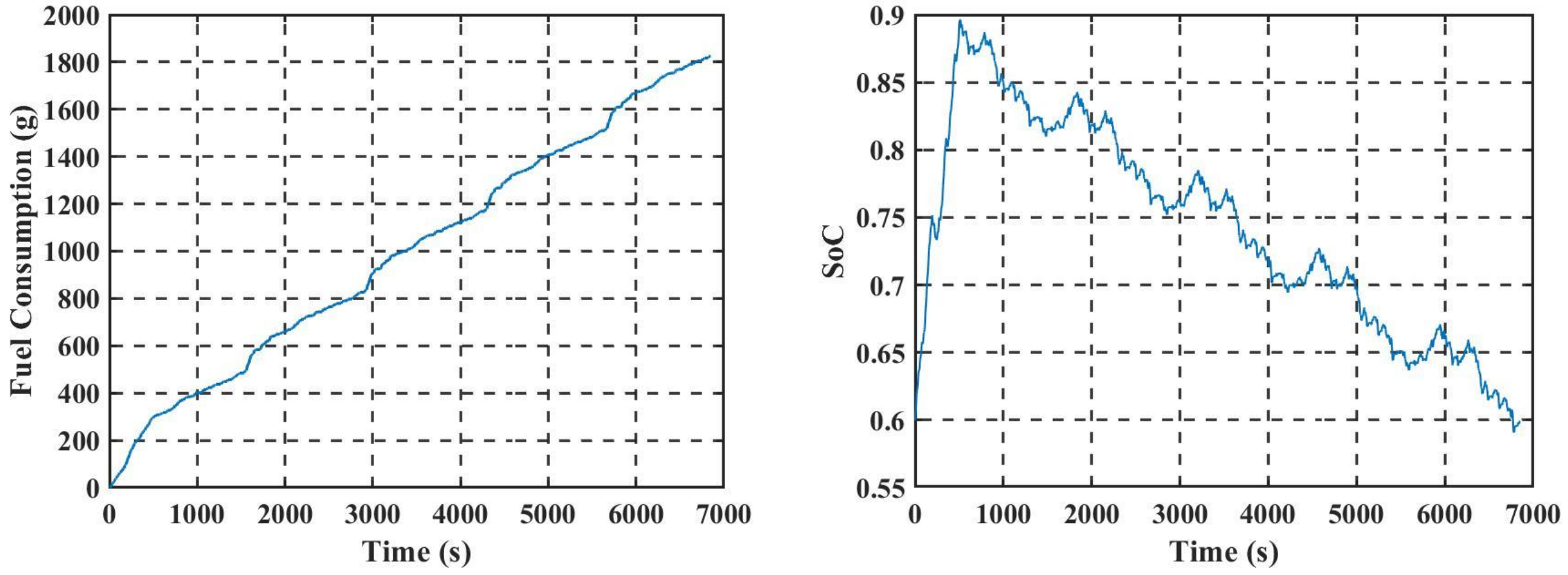

(a) Fuel consumption of engine. (b) SOC of battery.

Fig. 10. Simulation results of DP+ with 0.05Hz active signal injection.

In addition, the current profiles for different cases are analyzed in the frequency domain. As shown in Fig. 11(a), the maximum amplitude is close to 2A which means the richness of the baseline current signal is insufficient for battery parameters/states estimation. Applying the proposed DP+, the signals with the desired frequencies can be injected successfully, as shown in Figs. 11 (b) and (c).

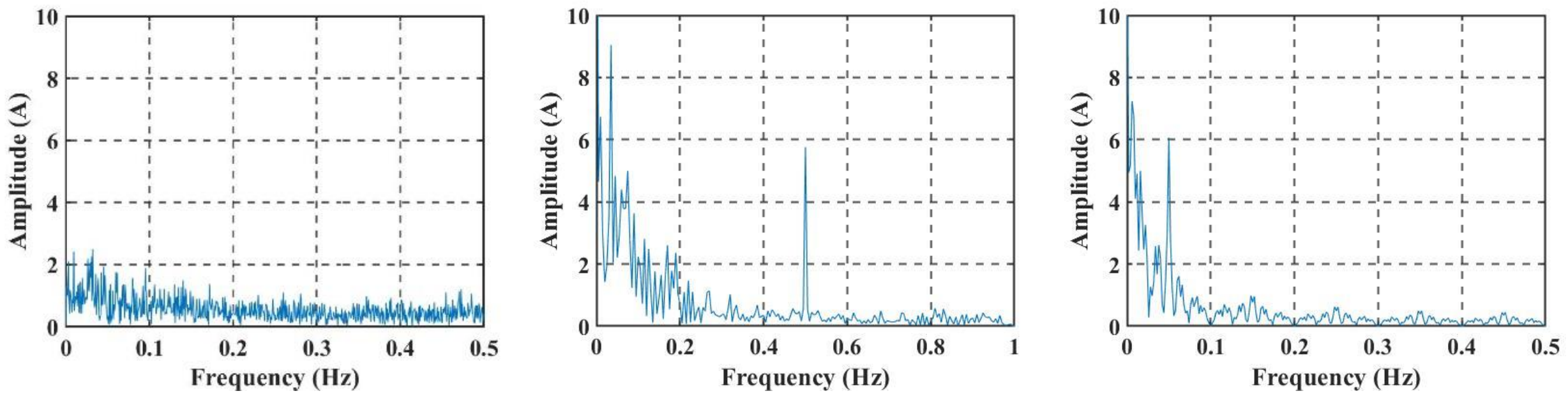

(a) DP-. (b) DP+ with 0.5Hz signal. (c) DP+ with 0.05Hz signal.

Fig. 11. Fast fourier transform of battery current.

### 5.3 Battery Parameter/State Estimation

After the simulations using the proposed DP+, the battery parameter/state identification is conducted using the current profiles. The sequential algorithm is adopted to estimate the ohmic resistance, parameters of the RC pair, battery capacity and SOC, based on the first-order equivalent circuit model. In order to simulate the real-life situation, white noise with an RMS value of 10mV is added into the voltage measurement. The initial guesses of the identified parameters and states are $[R_s(1)\ R_t(1)\ \tau(1)\ Q_b(1)\ SOC(1)] = [0.02\ 0.01\ 10\ 2\ 0.5]$. In Step #1, a first-order high-pass filter with 3dB bandwidth at 0.2Hz is chosen and EKF is applied to estimate the ohmic resistance of the battery. The estimated resistance can converge to the actual value quickly, as shown in Fig. 11 (a). In Step #2, a high-pass filter with 3dB bandwidth at 0.02Hz is selected and EKF is also adopted. The estimation process starts at 300s so that the initial SOC dynamics can be filtered. It is illustrated in Fig. 11(b) that the identified parameters can track the actual values accurately. In Step #3, SOH and SOC are estimated simultaneously based on the previous identified parameters. No excitation current needs to be injected in this step because the estimation of SOC is not affected by current frequency, while the identification of SOH prefers low current frequencies [31]. As shown in Fig. 11(c), the estimated values of capacity and SOC converge to the actual values in just one driving cycle. Therefore, it is verified that the battery parameters/states can be estimated accurately in a specific sequence using the sequential algorithm based on active signal injection.

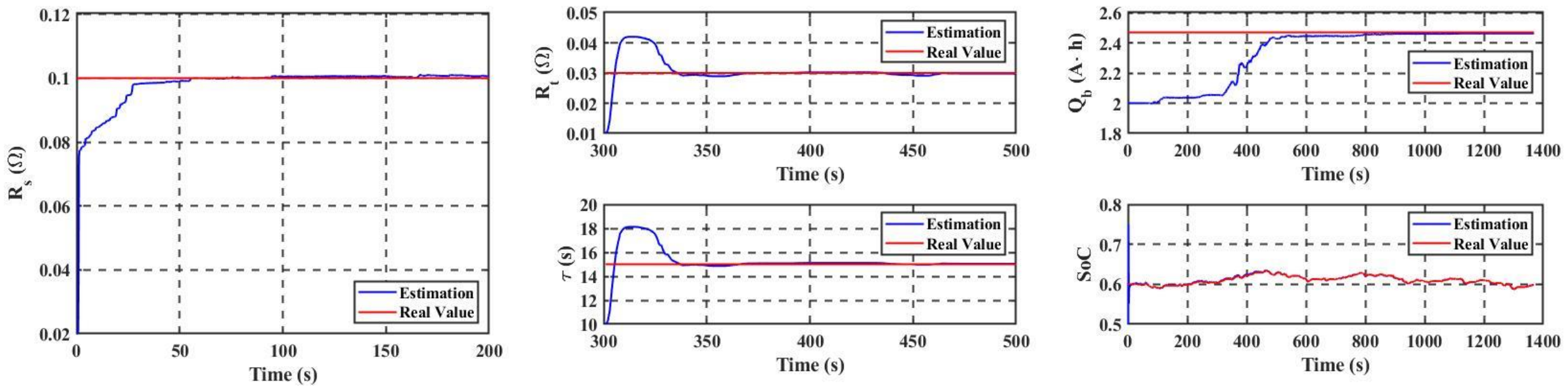

(a) Step #1. (b) Step #2. (c) Step #3.

Fig. 12. Estimation results of DP+: (a) ohmic resistance, (b) parameters of RC pair (c) capacity and SOC.

For comparison, the parameters and states of the battery are also identified concurrently using the current profile of the baseline DP- and the multi-scale DEKF. The initial conditions and measurement noise are set to be the same as those applied in the sequential algorithm. As shown in Fig. 13(a), the RMS of the ohmic resistance estimation increases by 100% when compared to the results using the proposed DP+ and the sequential algorithm. Although the identified $R_t$ converges to the actual value, it takes a longer time to do so and so the RMS error increases by 441%. In addition, the estimated $\tau$ and $Q_b$ can not track the actual values and there are significant static errors, as shown in Figs. 13 (b) and (c).

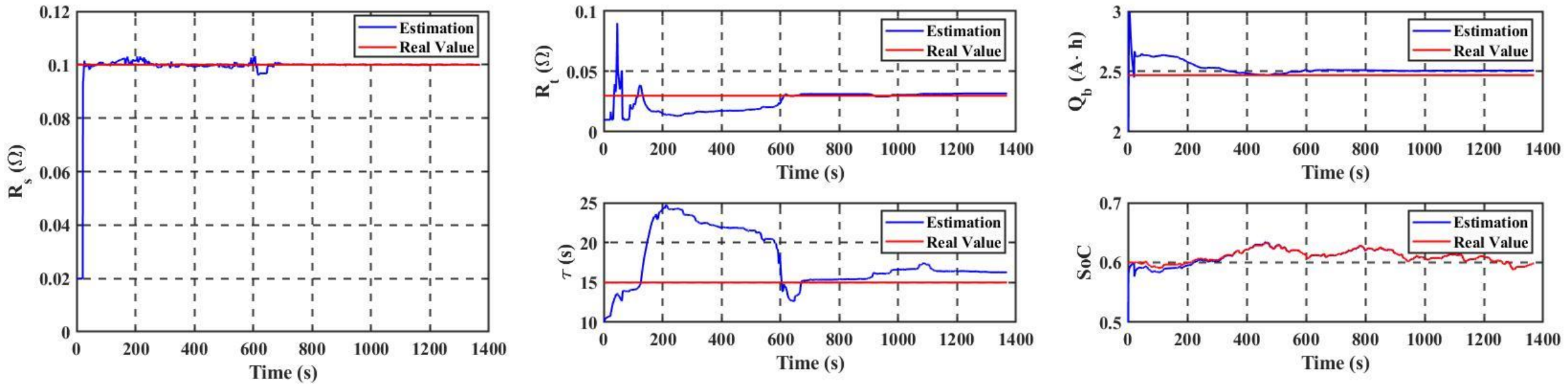

(a) Ohmic resistance. (b) Parameters of RC pair. (c) Capacity and SOC.

Fig. 13. Estimation results of DP- (parameters and states of battery are identified concurrently using DEKF).

## 6. Conclusion

Based on the over-actuated nature of the series HEV, SIC can be adopted in order to ensure the safe and efficient application of battery. DP is applied to optimize the fuel economy when active signals are injected for battery parameters/states estimation through the sequential algorithm. The method of how to successfully inject the signals is described (i.e., the battery current should have a constant value for every half period of excitation) and verified to be reasonable using the proposed DP. In addition, the tradeoff between fuel consumption and identification accuracy is exploited in order to provide a guideline for active signal injection (i.e., 6A is chosen to be the amplitude of the sinusoidal component of the battery current for this specific case). According to the simulation results, injecting active signals can improve identification accuracy of battery parameters/states by more than 100% and the increase of fuel consumption is slight, which is only 0.69% and 1.67% for different steps of the sequential algorithm. The proposed method of active signal injection can be further investigated in other cases (e.g. PHEV) using different PMSs (e.g. MPC) and the application range of SIC can be extended.